\documentclass[preprints,accept,moreauthors,pdftex]{Definitions/mdpi}

\usepackage{geometry}
\usepackage{graphicx}
\usepackage{xspace}
\usepackage{mathrsfs}
\usepackage{todonotes}
\reversemarginpar
\usepackage{changes}

\firstpage{1}
\pubvolume{1}
\issuenum{1}
\articlenumber{0}
\pubyear{2021}
\copyrightyear{2020}
\datereceived{}
\dateaccepted{}
\datepublished{}
\hreflink{https://doi.org/} 

\Title{Muon to positron conversion}

\TitleCitation{Muon to positron conversion}

\Author{MyeongJae Lee $^{1,\dagger}$\orcidA{} and Michael MacKenzie $^{2,\dagger}$\orcidB{}}
\address{%
$^{1}$ \quad Sungkyunkwan University, Suwon 16419, Republic of Korea; myeongjaelee@skku.edu\\
$^{2}$ \quad Northwestern University, Evanston, Illinois, 60208, USA; michaelmackenzie@u.northwesthern.edu}
\corres{Correspondence: michaelmackenzie@u.northwesthern.edu}

\firstnote{These authors contributed equally to this work.}

\abstract{
  Lepton flavor violation (LFV) has been discovered in the neutrino sector by neutrino oscillation experiments.
  The minimal extension of the Standard Model (SM) to include
  neutrino masses allows LFV in the charged sector (CLFV) at the loop-level, but
  at rates that are too small to be experimentally observed. Lepton number violation (LNV) is explicitly forbidden even in the
  minimally extended SM, so the observation of a LNV process would be unambiguous evidence of physics
  beyond the SM. The search for the LNV and CLFV process \mupprocess (referred to as $\MuToEp$ conversion) complements \znbb
  decay searches, and is sensitive to potential flavor effects in the neutrino mass generation mechanism.
  A theoretical motivation for $\MuToEp$ conversion is presented along with a review of the status of past $\MuToEp$ conversion
  experiments and future prospects.
  Special attention is paid to an uncertain and potentially dominant background for these searches, namely, radiative muon capture (RMC).
  The RMC high energy photon spectrum is theoretically understudied and existing measurements insufficiently constrain this portion
  of the spectrum, leading to potentially significant impacts on current and future $\MuToEp$ searches.
}

\newcommand{\kevc}      {keV$\!$/$\!$c}
\newcommand{\mevc}      {MeV$\!$/$\!$c}
\newcommand{\ra}        {\mbox{$\rightarrow$}}
\newcommand{\MuToEp}    {\mbox{$\mu^- \ra ~e^+$}}
\newcommand{\MuToEm}    {\mbox{$\mu^- \ra ~e^-$}}

\newcommand{\mupconversion}{$\mu^{-} \to e^{+}$\xspace}
\newcommand{\mueconversion}{$\mu^{-} \to e^{-}$\xspace}
\newcommand{\znbb}{$0\nu\beta\beta$\xspace}
\newcommand{\mupprocess}{$\mu^- + N(A,Z) \to e^+ + N'(A,Z-2)$\xspace}

\reversemarginpar

\begin{document}

\section{Introduction}

The incoherent conversion of a negative muon into a positron in a muonic atom, \mupprocess (referred to as \mupconversion hereafter),
is an exotic process that is both
lepton flavor violating (LFV) {\it and}
lepton number violating (LNV) with a change in lepton number by two units ($\Delta L = 2$).
The conservation of charged lepton number and flavor has been very well experimentally established. The symmetry corresponding
to the conservation of lepton flavor is broken in the Standard Model (SM) by the introduction of neutrino masses, and the symmetry
corresponding to the conservation of lepton number can additionally be broken by the introduction of new interactions and particles.
In the last few decades, LFV decays of the muon have been studied experimentally through three processes, each forbidden before
the introduction of neutrino masses:
$\mu^+ \to e^+ \gamma$, $\mu^+ \to e^+ e^- e^+$, and $\mu^- + N \to e^- + N$ (referred to as \mueconversion).
Experiments searching for \mueconversion can also typically search for \mupconversion, and so are able to
investigate both LFV and LNV in the muon sector.
The most recent result on muon LFV processes comes from the MEG experiment searching for
$\mu^+ \to e^+ \gamma$, 
reporting
$B(\mu^+ \to e^+ \gamma) < 4.2 \times 10^{-13}$
at 90\% confidence level (CL)\,\cite{Baldini2016}.
The upgrade of the MEG experiment to improve the
experimental sensitivity by
an order of magnitude is in progress\,\cite{sym13091591}.
The new Mu3e\,\cite{ARNDT2021165679} at PSI will search for $\mu^+ \to e^+ e^- e^+$,
improving the sensitivity by three orders of magnitude in Phase I.
New results for \mueconversion with 100 - 10,000 times better sensitivity are expected
by the end of the decade from the COMET\,\cite{10.1093/ptep/ptz125} and Mu2e\,\cite{bartoszek2015mu2e} experiments.

The theoretical branching ratio expectations for muon LFV decays in the SM minimally extended to include neutrino masses
are extremely small, below $10^{-50}$ for all three searched for processes.
The rate is suppressed by $((\Delta m_\nu^2)/M_W^2)^2$ where $m_\nu$ is the neutrino mass and
$M_W$ is the W-boson mass\,\cite{MarcianoWilliamJ2008CLFV},
and also due to the GIM (Glashow–Iliopoulos–Maiani) mechanism.
Therefore, observation of any charged LFV (CLFV) process is direct evidence of physics beyond the SM.

\begin{figure}[H]
  \centering
  \includegraphics[width=0.5 \textwidth]{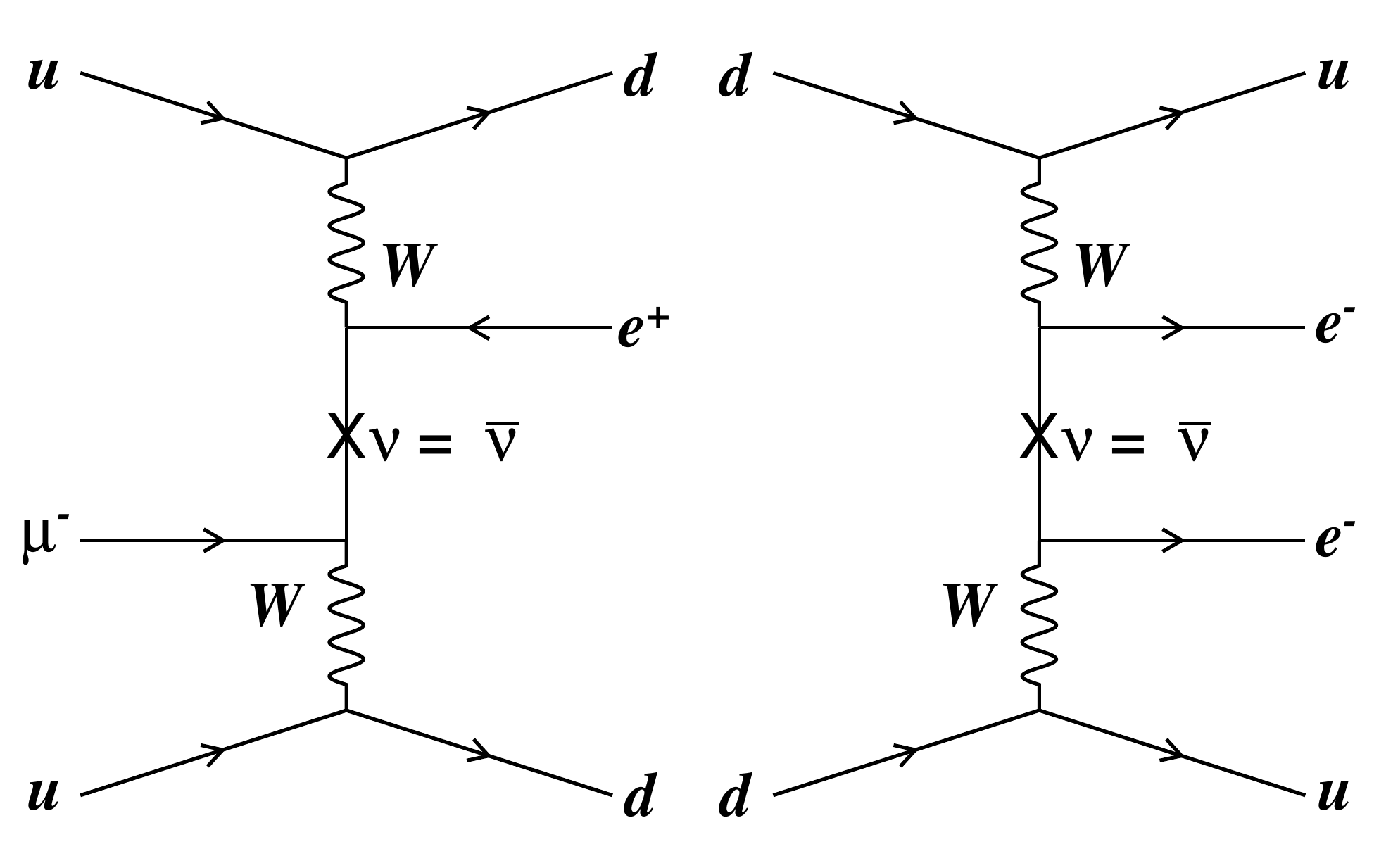}
  \caption{The tree level diagram of \mupconversion (left) and \znbb (right) in a Majorana neutrino model.
    \label{fig_mup_diagram}}
\end{figure}

In the LNV case, such as \mupconversion, the process is not allowed as a perturbative process.
The minimal extension of the SM with a Majorana neutrino allows for the
famous LNV process, neutrinoless double-beta decay (\znbb), to occur.
In fact, the Feynman diagrams for \mupconversion and \znbb mediated by a Majorana neutrino
are very similar, as depicted in Figure\,\ref{fig_mup_diagram}, except that
\mupconversion also involves a change of lepton flavor.
The \mupconversion and \znbb processes are complimentary in the sense that \znbb involves same-flavor (or ``flavor-diagonal'')
transitions while \mupconversion involves a different flavor (of ``flavor-off-diagonal'') transition. In some theories for physics
beyond the SM, flavor-diagonal transitions are suppressed and observable signals of new physics are associated with different-flavor
processes.

Benefiting from progress in detector technology and the feasibility of obtaining a large amount
of target material,
models involving LNV in general are more easily observable at \znbb experiments,
such as KamLAND-Zen\,\cite{Gando_2020}, than in \mupconversion searches.
This is true for the case that a light Majorana neutrino mediates the LNV,
where \znbb would be more easily measurable than \mupconversion.
The effect on the \znbb and \mupconversion processes in a specific new physics model, for example, a heavy sterile neutrino model,
is not very well studied.
The ``Black box theorem''\,\cite{PhysRevD.25.2951,NIEVES1984375,TAKASUGI1984372,HIRSCH2006106}
relates the Majorana neutrino mass to the amplitude of \znbb and any $\Delta L = 2$ processes even in the presence of new physics,
showing that the occurrence of any $\Delta L = 2$ process, including \mupconversion, implies a nonzero Majorana neutrino mass.
An enhancement or suppression of the rate of \mupconversion over \znbb was suggested by some new physics
models\,\cite{PhysRevD.95.115010,GEIB2017157,PhysRevD.95.055009,PhysRevD.75.053004,King2014,Pritimita2016,PhysRevLett.93.231802}.
In addition, some theories explain the LNV of \znbb without a Majorana neutrino exchange,
but instead using a new mediator particle from a supersymmetric theory or a Majoron: see the discussion by
Engel and Men\'endez\,\cite{Engel_2017}.
These models discussed by Engel and Men\'endez can also be used to describe \mupconversion.
Therefore, from one perspective, \mupconversion is a LNV search mediated by a Majorana neutrino similar
to \znbb experiments, and from another perspective,
it is a new physics search regarding the flavor effect on the neutrino
mixing in the framework of the theory beyond the SM.
While there are other LNV searches such as $\tau^- \to e^+ \pi^- \pi^-$\,\cite{MIYAZAKI2013346}
or $K^\pm \to \pi^\mp \mu^\pm \mu^\pm$\,\cite{2019134794}, those are not similar processes
to \znbb or \mupconversion, 
so their direct comparison with \mupconversion or \znbb is very difficult. 
A few very recent studies have suggested comparing the \znbb results with the collider experimental results for the $t$-channel process of same-sign W-boson scattering\,\cite{PhysRevD.103.115014,PhysRevD.103.055005,Cirigliano2021}.

From the experimental point of view, \mupconversion is attracting more interest as the future
\mueconversion experiments progress. The COMET\,\cite{10.1093/ptep/ptz125} and
Mu2e\,\cite{bartoszek2015mu2e} experiments will search for
\mueconversion 
with an experimental sensitivity
down to $10^{-17}$.
Although the experimental designs were not optimized for \mupconversion,
these experiments can significantly improve the sensitivity reach of \mupconversion
owing to  the considerable increase in the number of muons.
The current best experimental limits on \mupconversion were obtained by the SINDRUM II
experiment with titanium nuclei:
$B(\mu^{-}+{~\rm Ti} \to e^{+}+{~\rm Ca})< 1.7 \times 10^{-12}$
at 90\% CL to the ground state of calcium\,\cite{1998334}.

The purpose of this article is to emphasize the importance of the
search for \mupconversion and assess its feasibility in future \mueconversion experiments.
Section \ref{section2} reviews the theoretical estimation of the \mupconversion rate in connection with the \znbb search.
It also describes the past experimental results for \mupconversion.
Section \ref{section3} introduces the future \mueconversion experiments where \mupconversion can be studied.
A major background component is analyzed and a mitigation strategy for suppressing the background in order to improve
the signal sensitivity of \mupconversion searches is also described.

\section{Theories and past results}
\label{section2}

\subsection{Estimation of the $\mu^- \to e^+$ rate in the extended SM with a Majorana neutrino}

A generic model of neutrino masses includes at least three light, left-handed neutrinos, with neutrino flavor oscillations
described by the Pontecorvo–Maki–Nakagawa–Sakata (PMNS) matrix\,\cite{Pontecorvo:1957qd,Pontecorvo:1957cp,10.1143/PTP.28.870}.
A right-handed neutrino, or any other hypothetical neutrino, may have significantly larger mass
than the light neutrinos, as is expected, for example, from the Seesaw
Mechanism\,\cite{MINKOWSKI1977421,Gell-Mann:1979vob,10.1143/PTP.64.1103,10.1007/978-1-4684-7197-7_15,PhysRevLett.44.912,PhysRevD.22.2227}.
In the minimal extension of the SM with a Majorana neutrino, the \mupconversion process is allowed
through a light or heavy Majorana neutrino exchange, as depicted in Figure\,\ref{fig_mup_diagram}.
But, as described before, it is also possible that other new particles beyond the SM
or new interactions between quarks and leptons may mediate the interaction.

Assuming the light or heavy Majorana neutrino exchange interaction,
the leading order \mupconversion matrix element when the initial and final states of nuclei
are both ground states can be written as\,\cite{PhysRevC.70.065501}:
\begin{eqnarray}
  \mathscr{M}_{fi} & = & -i \left(  \frac{G_F}{\sqrt{2}} \right) ^2 \frac{1}{\left( 2\pi\right) ^{3/2}} \frac {1}{\sqrt{4E_{\mu^-} E_{e^+}}}
  \bar{v} ( k_{e^+} )  ( 1+\gamma_5 )  u( k_{\mu^-})  \nonumber\\
  & & \times \frac{m_e g_A ^2}{2\pi R} \left[ \frac{\langle m_\nu \rangle_{\mu e}}{m_e} \mathcal{M}_\nu + \langle M_N ^{-1} \rangle_{\mu e} m_p
    \mathcal{M}_N \right]
  \times 2 \pi \delta \left( E_{\mu^-} + E_i - E_f - E_{e^+} \right) \label{eq:mup_rate}\\
 & \propto & \left\{ \begin{array}{ll}
 \langle m_\nu \rangle_{\mu e} \mathcal{M}_\nu  & {\rm (light~neutrino)}  \\
 \langle M_N ^{-1} \rangle_{\mu e}  \mathcal{M}_N & {\rm (heavy~neutrino)} \end{array} \right. \nonumber
\end{eqnarray}
In this equation,
$G_F$ is the Fermi constant,
$m_e$ is the electron mass,
$g_A$ is the weak axial coupling constant, and
$R$ is the nuclear radius.
$E_{(\mu^-, e^+, f, i)}$ represent the energy of the muon, positron, final nuclear
ground state, and initial nuclear
ground state, respectively.
The effective neutrino masses are defined as:

\begin{equation}
\langle m\rangle_{\alpha \beta} = \sum_k U_{\alpha k} U_{\beta k} m_k ~,
\end{equation}

\noindent
for the light neutrino, and

\begin{equation}
\langle M^{-1} _N \rangle_{\alpha \beta} = \sum_k \frac{U_{\alpha k} U_{\beta l}}{M_k}~,
\end{equation}

\noindent
for the heavy neutrino, where
$\alpha$ and $\beta$ are flavor indexes, and $m_k$ and $M_k$ represent the light neutrino mass and the heavy neutrino mass, respectively.
$U$ is the neutrino mixing matrix, or PMNS matrix for the light neutrino case.
The effective neutrino mass matrix can be calculated from the current estimation of the neutrino
masses and mixing angles, in the case of the light neutrino exchange model.
For example, from a cosmological observation, the sum of three light neutrino masses is less than
0.42\,eV, which translates to $\langle m_\nu \rangle_{l_1 l_2} < 0.14\,{\rm eV}$\,\cite{PhysRevD.71.113014}.

Another important term in Equation\,\ref{eq:mup_rate} is $\mathcal{M}_{\nu / N}$, representing
the nuclear matrix element (NME) for the light and heavy neutrino, respectively. This is a transition
probability matrix of the nucleon from the initial state to the final state:

\begin{equation}
\mathcal{M}_i \propto \int dq \sum_n \langle f | J^\mu (x) | n \rangle \langle n | J^\nu (y) | i \rangle  = \mathcal{M}_i ^{(GT)} + \mathcal{M}_i ^{(T)} + \mathcal{M}_i ^{(F)}~,
\end{equation}

\noindent
where $n$ is an intermediate nuclear state and $q$ and $J$ represent the momentum transfer and hadronic
current, respectively.
Theoretically, the matrix element is the sum of three components:
the axial vector ($\mathcal{M}_i ^{(GT)}$, Gamow-Teller term),
tensor ($\mathcal{M}_i ^{(T)}$), and vector ($\mathcal{M}_i ^{(F)}$, Fermi term)
terms, where the axial vector NME is the dominant one\,\cite{Engel_2017,10.3389/fphy.2019.00030}.
Because of uncertainties in the nucleus models, and difficulty of calculating
many body dynamics, the NME calculation usually needs to be approximated.
Approximation methods of NME calculations used for \znbb experiments,
such as the interacting shell model (ISM), quasiparticle random-phase approximation (QRPA),
interacting boson model (IBM), projected Hartree-Fock-Bogoliubov model (PHFB), and
energy density functional method (EDF),
are reviewed by Vergados et al.\,\cite{Vergados_2012} and Engel and Men\'endez\,\cite{Engel_2017}.
Engel and Men\'endez\,\cite{Engel_2017} 
also show the NME calculations for \znbb in different materials and with different approximation methods differ by up to
a factor of three.
This shows the importance of understanding the NME for the \znbb process, and therefore for the similar \mupconversion process,
in the minimal extension of the SM with a Majorana neutrino.

Using this formulation, the theoretical estimation of the \mupconversion rate was obtained
by Domin et al.\,\cite{PhysRevC.70.065501}:

\begin{eqnarray}
\mathcal{R}^{\mu^- e^+} & \equiv & \frac{\Gamma(\mu^- + N(A,Z) \to e^+ + N'(A,Z-2))}{\Gamma(\mu^- + N(A,Z) \to {\rm (All~muon~captures)})} \\
 & = & 2.6 \times 10^{-22}  \times   \left\{ \begin{array}{ll}
 \left| \langle m_\nu     \rangle_{\mu e} / m_e \right|^2 \left| \mathcal{M}_\nu \right|^2 & {\rm (light~neutrino)}  \\
 \left| \langle M_N ^{-1} \rangle_{\mu e} m_p   \right|^2 \left| \mathcal{M}_N   \right|^2 & {\rm (heavy~neutrino)}~ \end{array} \right\} \nonumber
\end{eqnarray}

\noindent
Applying the effective neutrino mass obtained from the experimental data on the neutrinos
masses and their mixing,
and the NME for titanium from the QRPA method, $\mathcal{R}^{\mu^- e^+}({\rm Ti})$ is:
\begin{equation*} \begin{array}{rl}
 (0.008 - 1.7) \times 10^{-41} & {\rm for~a~light~neutrino,~normal~neutrino~mass~hierarchy,} \\
 (0.05 - 6.7) \times 10^{-40} & {\rm for~a~light~neutrino,~inverted~neutrino~mass~hierarchy,~and} \\
 \le 3.8 \times 10^{-24} &  {\rm for~a~heavy~neutrino.} \end{array}
\end{equation*}

\noindent
While the estimated \mupconversion rate is much higher in the  heavy Majorana neutrino case
than the light neutrino one, it is far smaller than the
feasible experimental reach.

\begin{figure}[tb]
  \centering
  \includegraphics[width=0.6 \textwidth]{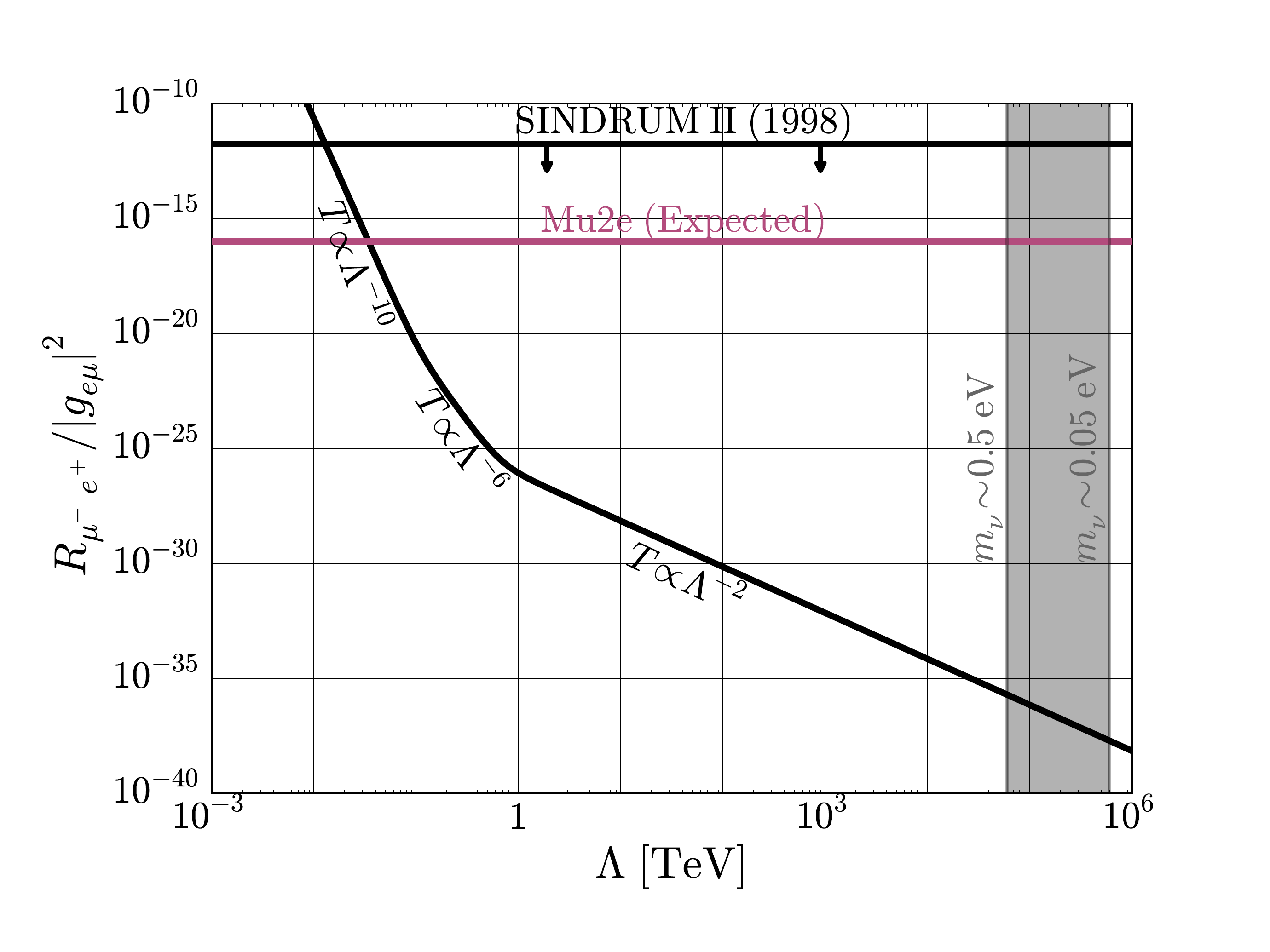}
  \caption{The estimation of the \mupconversion rate, as a function of the new physics scale parameter $\Lambda$,
    taken from Ref. \cite{PhysRevD.95.115010}.
  }
  \label{fig_ode_estimation}
\end{figure}

Any  $\Delta L = 2$ interaction other than the tree-level interaction depicted
in Figure\,\ref{fig_mup_diagram}  may lead to \mupconversion. This was studied by
Berryman et al.\,\cite{PhysRevD.95.115010} by using an effective operator description
and normalizing the estimation of the Majorana neutrino exchange case (including one- and two-loop corrections)
with the above tree-level calculation.
The conversion rate was estimated according to the
new physics energy scale
($\Lambda$),
shown in Figure\,\ref{fig_ode_estimation}.
From the expected sensitivity of Mu2e, the new physics energy scale accessible from a
\mupconversion measurement is comparably low, around 40\,GeV,
compared to the new physics scale accessible from a \mueconversion measurement at Mu2e of
$\mathcal{O}(10^4)$\,TeV.
For all possible $\Delta L = 2$ interactions leading to \mupconversion and \znbb, the new physics scale reach
of \mupconversion searches
is a few orders of magnitude smaller than \znbb experiments.
According to Berryman et al.\,\cite{PhysRevD.95.115010},
the observation of \mupconversion 
would imply that:
first, the neutrino is a Majorana fermion;
second, flavor effects suppress \znbb while enhancing \mupconversion; 
and third, a combination of complex interactions other than the tree-level interaction is responsible for
the physics of nonzero neutrino mass.

\subsection{Past $\mu^- \to e^+$  experiments}

The results from past \mupconversion experiments are listed in Table\,\ref{table_past_results}\,\cite{10.1093/ptep/ptaa104}.
The experimental techniques were similar for most all of these searches: a muon beam interacted with a nuclear target, where muons
quickly fell to a 1s orbit in a target atom. The outgoing positron spectrum from the nuclear target was measured in
a tracking detector, searching for the \mupconversion signal.
The experiment using ${}^{127}{\rm I}$ as the nuclear target\,\cite{ABELA1980318} used a radiochemical method: a muon beam interacted
with a NaI target, and then the target was chemically treated to extract ${}^{127}{\rm Sb}$
or ${}^{127}{\rm Te}$, where the decay rate of ${}^{127}{\rm Te}$ was measured to detect an excess of ${}^{127}{\rm Sb}$.

\begin{figure}[tb]
\widefigure
\includegraphics[width=0.49 \linewidth]{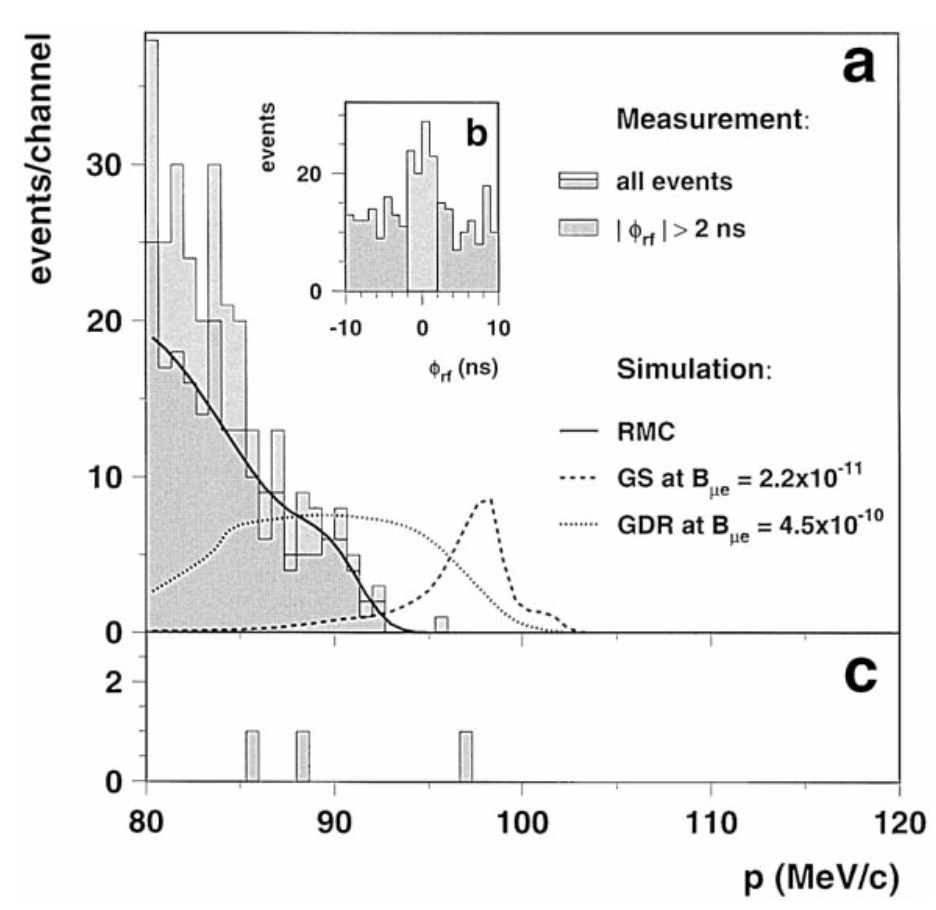}
\includegraphics[width=0.49 \linewidth]{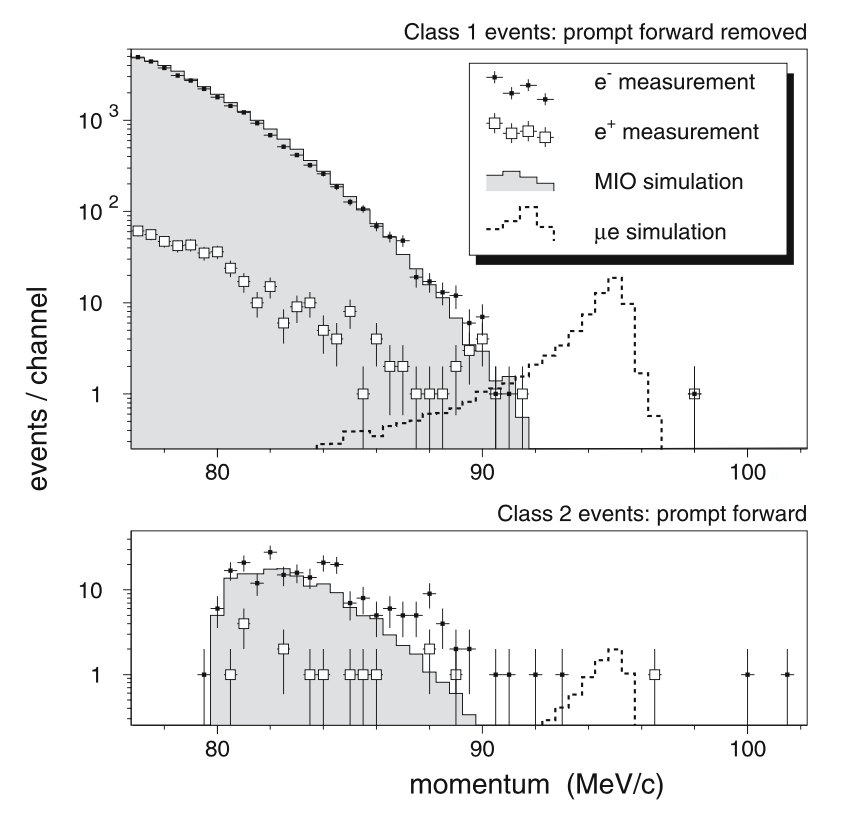}
\caption{The energy spectrum of the signal estimation and backgrounds of \mupconversion (left) and \mueconversion (right) 
from the SINDRUM II experiment\,\cite{1998334,sindrum_ii:Bertl2006}.
For \mupconversion, the and lower plots are the data when the muon beam is on and off, respectively.
For \mueconversion, the and lower plots are the data with the radiative pion background candidates removed
and enhanced, respectively.
\label{fig:sindrumII_spectrums}}
\end{figure}

Figure\,\ref{fig:sindrumII_spectrums} shows the most recent
\mupconversion experimental search\,\cite{1998334} and the comparison to the most recent \mueconversion
search\,\cite{sindrum_ii:Bertl2006}, both from SINDRUM II.
For \mupconversion, the nucleus in the final state can be either in the
ground state or an excited state.
In the case of the transition to the ground state,
the signal positron is mono-energetic with an energy ($E_{\mu^-e^+}$) given by:

\begin{equation}
E_{\mu^-e^+} = m_\mu +M(A,Z) - M(A,Z-2) - B_\mu - E_{\rm{recoil}}\,,
\label{eq:mup_energy}
\end{equation}

\noindent
where $m_{\mu}$ is the muon mass, $M(A,Z)$ is the mass of the nucleus $N(A,Z)$, $B_{\mu}$
is the binding energy of the muonic atom, and $E_{\rm recoil}$ is the recoil energy of the outgoing nucleus.
Similarly to photo-nuclear reactions, the \mupconversion process
may leave the target nuclei in an excited state,
a Giant Dipole Resonance (GDR), which is a collective vibration of protons against neutrons with a dipole
spatial pattern\,\cite{doi:10.1146/annurev.ns.36.120186.002553}.
For example, a GDR with a 20\,MeV width was assumed in Ref.\,\cite{1998334},
resulting in a much wider energy distribution for the signal positron, as can be seen
in Figure\,\ref{fig:sindrumII_spectrums}.
The past experimental results in Table\,\ref{table_past_results} are separately reported
for the final ground state and GDR state of the nucleus.
Table\,\ref{table_nuclei_list} lists the $E_{\mu^-e^+}$ of the ground state transition for some selected nuclei
used by or considered for \mupconversion experiments.

\end{paracol}
\begin{specialtable}[bt]
\widetable
\caption{
The past \mupconversion experimental results\,\cite{10.1093/ptep/ptaa104}. The limits are quoted at
90\,\% CL, and
$\rm N_{capture}$ is the number of muon captures in the experiments.
GS and GDR represent 
ground
state and giant dipole resonance excitation of the final nucleus status.
While the natural abundance of ${}^{32}{\rm S}$ and
${}^{127}{\rm I}$ among their isotopes are almost 100\,\%, 
Ti and Cu are not, which is the reason for not specifying the mass of Ti and Cu.
\label{table_past_results}}
\begin{tabular}{cccccccc}
\toprule
\textbf{Nuclei} & \textbf{Upper limit} & $\mathbf{N_{capture}}$ & \textbf{Year}  & \textbf{Experiment} & \textbf{Detector} & \textbf{GS/GDR} & \textbf{Reference}\\

\midrule

\multirow{5}{*}{Ti}
 & $3.6 \times 10^{-11}$ & \multirow{2}{*}{\Big\} $2.5 \times 10^{13}$} & \multirow{2}{*}{1998} & \multirow{2}{*}{SINDRUM II} & \multirow{2}{*}{Drift chamber} & GDR  & \multirow{2}{*}{\cite{1998334, sindrum_ii:Kaulard1997_Thesis}} \\
 & $1.7 \times 10^{-12}$ &  &  &  &  & GS & \\
 & $8.9 \times 10^{-11}$ & \multirow{2}{*}{\Big\} $4.9 \times 10^{12}$} & \multirow{2}{*}{1993} & \multirow{2}{*}{SINDRUM II} & \multirow{2}{*}{Drift chamber} & GDR & \multirow{2}{*}{\cite{DOHMEN1993631}} \\
 & $4.3 \times 10^{-12}$ & &  & & & GS &  \\
 & $1.7 \times 10^{-10}$ & $9 \times 10^{12}$   & 1988 & (TRIUMF) & TPC & GDR & \cite{PhysRevD.38.2102}  \\

\midrule

\multirow{2}{*}{${}^{32}{\rm S}$}
 & $9 \times 10^{-10}$ &  $6.7 \times 10^{11}$ & 1980 &  SIN & Streamer chamber & GDR & \cite{Badertscher:1980bt} \\
 & $1.5 \times 10^{-9}$ & $1.2 \times 10^{11}$ & 1978 & SIN & Streamer chamber & GDR & \cite{1979434, BADERTSCHER1978371} \\

\midrule

\multirow{2}{*}{Cu}
 & $2.6 \times 10^{-8}$ &                   & 1972 &        & Spark chamber & GS/GDR & \cite{PhysRevLett.28.1469}  \\
 & $2.2 \times 10^{-7}$ & $2.2 \times 10^9$ & 1962 & (CERN) & Spark chamber &   & \cite{NC.26.261}            \\

\midrule

${}^{127}{\rm I}$ & $3 \times 10^{-10}$ & $2.1 \times 10^{12}$ & 1980 & & Radiochemical & GS  & \cite{ABELA1980318} \\

\bottomrule
\end{tabular}
\end{specialtable}
\begin{paracol}{2}
\switchcolumn

\end{paracol}
\begin{paracol}{2}
\switchcolumn

\section{Future $\MuToEp$ experimental searches}
\label{section3}

\subsection{Upcoming experimental prospects}

The COMET Phase-I\,\cite{10.1093/ptep/ptz125} and Mu2e\,\cite{bartoszek2015mu2e} experiments will have unprecedented sensitivity
to $\MuToEp$ using aluminum as their nuclear targets, with single event sensitivities (SES)
on the order of $10^{-15}$ and $10^{-17}$ respectively in the $\MuToEm$ conversion search.
These experiments are expected to have similar sensitivities in the $\MuToEp$ channel.
After COMET Phase-I, the detector will be upgraded for COMET Phase-II which will have an SES on the order of $10^{-17}$\,\cite{10.1093/ptep/ptz125}.
COMET Phase-II will not be able to search for positron and electron signals simultaneously, so a \mupconversion search would be
a special run of the \mueconversion experiment and not necessarily reach the same muon statistics as estimated for the electron
channel.
There has also been an expression of interest in upgrading the Mu2e detector, called Mu2e-II, with an SES on the order of
$10^{-18}$\,\cite{Mu2e_II_abusalma2018expression}.

Both COMET Phase-I and Mu2e use similar principles to search for the $\MuToEm$ and $\MuToEp$ processes:
(1) a pulsed proton beam is used to produce
pions and muons in a production target within a superconducting solenoid; (2) the low momentum pions and muons are guided by the
superconducting production solenoid field to a curved transport superconducting solenoid, which has an acceptance designed to
eliminate high momentum particles; (3) an off-axis collimator selects the desired charge of the beam, utilizing the sign-dependent
drift of the beam along the curved beamline; (4) the muons from production and pion decays along the beamline are stopped in a
nuclear target within a superconducting solenoid containing the cylindrical detector elements; (5) non-stopped beam particles continue
through the solenoid,
passing through a central axis hole in the detector elements; (6) after a time sufficient for nearly all pions to either decay or be
captured on the nuclear target, signal candidates are reconstructed in the detectors (where the central holes blind them to
the high intensity, low momentum muon decay backgrounds).
COMET Phase-II is similar to Phase-I, except after the stopping target there is a second curved transport solenoid followed by the
detectors,
where the second transport solenoid provides the momentum filtering, eliminating the need for a central hole in the detectors.
This second transport solenoid is also what prevents the simultaneous search for $\MuToEm$ and $\MuToEp$.
The pulsed proton beam significantly reduces beam related backgrounds during the signal measurement period
by delaying the signal
measurement period until after the beam products arrive. This
time delay method requires that the lifetime of the muon in the nuclear target (see Table\,\ref{table_nuclei_list}) is large enough that there are
a sufficient number of muon captures/decays in the target during the signal search period. Due to this restriction, nuclear target materials with
$Z \gtrsim 40$ cannot be easily studied at experiments like COMET and Mu2e.

\subsection{Background consideration}

A potentially dominant physics background to \mupconversion comes from radiative muon capture (RMC),

\begin{equation*}
\mu^- + N(A,Z) \to \nu_\mu + N(A,Z-1) + \gamma~,
\end{equation*}

\noindent
followed by photon pair production, $\gamma \to e^+e^-$,
where the $e^{+}$ in the pair is misidentified as a signal.
The photon energy spectrum of RMC, in particular in the endpoint energy region,
is poorly known experimentally\,\cite{RMC_1999_PhysRevC.59.2853}.
However, the maximum allowed endpoint energy ($E^{\rm{end}}_{\rm{RMC}}$) can be kinematically determined:

\begin{equation}
E^{\rm{end}}_{\rm{RMC}} = m_\mu + M(A,Z) - M(A,Z-1) - B_\mu - E_{\rm{recoil}}\,.
\label{eq:rmc_endpoint}
\end{equation}

\noindent
The endpoint energies of some relevant nuclei are shown in Table\,\ref{table_nuclei_list}.
The nucleus in the final state may not be $N(A,Z-1)$ once nucleon emission occurs.
In this case, the RMC endpoint energy is smaller than that of Equation\,\ref{eq:rmc_endpoint}.
It is also possible that the RMC endpoint energy could be much smaller depending on the spin state
of the $N(A,Z)$ and $N(A,Z-1)$ nuclei.
In order to account for the RMC background, it is necessary to measure the RMC photon spectrum,
either at the \mueconversion experiment or in a dedicated experiment, such as the AlCap experiment\,\cite{Edmonds:2018agn}.
In addition, the kinematic separation of the RMC background and the \mupconversion positron
can be improved in future experiments such as COMET Phase-II or Mu2e-II by choosing a target material
where the endpoint energy of RMC is smaller than the \mupconversion positron energy,
$E^{\rm{end}}_{\rm{RMC}} < E_{\mu^-e^+}$\,\cite{PhysRevD.96.075027},
therefore:

\begin{equation}
M(A,Z-2) < M(A,Z-1)\,.
\label{eq:rmc_mass_requirement}
\end{equation}

Radiative pion capture (RPC),
\begin{eqnarray*}
\pi^- + N(A,Z) & \to & N(A,Z-1) + \gamma~~ {\rm and} \\
\pi^+ + N(A,Z) & \to & N(A,Z+1) + \gamma~,
\end{eqnarray*}
followed by $\gamma \to e^+ e^-$ is another background source,
when the converted positron is misidentified as a signal.
The RPC background can be controlled in \mupconversion and \mueconversion experiments by suppressing
the pion contamination in the muon beam. 
The pion contamination can be suppressed by using a beamline sufficiently long where the pions decay to muons
before reaching the nuclear target.
This was true for the previous \mueconversion measurements, where the $\pi / \mu$ ratio was about $10^{-7}$
for the SINDRUM II experiment for example, allowing the use of a continuous muon beam with continuous data acquisition\,\cite{1998334}.
The COMET and Mu2e \mueconversion experiments are targeting a factor of 100 to 10,000 improvement on the
upper limits for the \mueconversion process in the absence of a signal. To suppress the effects of pions and other
non-muon beam particles contaminating the beam, they are adopting a pulsed proton beam
with a high inter-beam pulse primary particle suppression and a delayed data acquisition timing window
technique\,\cite{bartoszek2015mu2e,10.1093/ptep/ptz125}.

Antiprotons contaminating the muon beam can annihilate in the muon stopping target and create high energy background
$e^\pm$ tracks, as well as introduce delayed pion background due to interactions along the muon beamline.
Along with the pulsed beam technique, antiproton absorbers can be placed in the muon beamline to
suppress the antiproton background.

Another large background arises from cosmic-ray induced events.
In general, the cosmic-ray induced background is similar for the \mueconversion and \mupconversion searches. Therefore,
the consideration of the detector design to suppress the cosmic-ray background for the \mueconversion search is also
applicable for the \mupconversion search.
The COMET and Mu2e experiments are developing cosmic-ray veto detector systems to detect charged cosmic-ray particles
entering the detectors and veto signal candidates coincident with these detected cosmic-rays.
Cosmic-rays can be reconstructed in the trackers at COMET Phase-I and Mu2e to constrain
the cosmic-ray veto detector systems' efficiencies, and in the case of COMET Phase-I where the tracking detector
covers the muon stopping target, cosmic-rays that interact with the stopping target and produce background tracks
can be fully reconstructed\,\cite{phd_tswong}.

Muon decay-in-orbit (DIO), which is one of the major background sources for \mueconversion,
is also a background source in the \mupconversion search due to photons generated in DIO electron interactions pair-creating
a positron in the detector material.
It is not an important background at the SES of the planned COMET and Mu2e experiments as long as the charge identification
is sufficient to suppress DIO electrons being reconstructed as positron events.
However, it should be noted that the charge identification will never be perfect, especially in the case that an electron or
positron is generated downstream of the detector system and propagates back to the muon stopping target, where only the tracker hit
timing and, for Mu2e and COMET Phase-II, the calorimeter information can distinguish the particle trajectory.

\subsection{RMC status}

Both COMET and Mu2e will have far greater sensitivity to the high momentum positron spectrum from RMC than previous
muon conversion and RMC measurement experiments.
The background to the \mupconversion search arises from highly asymmetric RMC photon conversions, either from on-shell photons
converting in the detector
material or off-shell photons internally converting, and the background strongly depends on the nuclear target.
The internal conversion spectrum for RMC has never been measured, though the internal conversion spectrum approximation
from Kroll and Wada\,\cite{RPC_1955_Kroll_Wada_PhysRev.98.1355}
(with corrections published by Joseph\,\cite{RPC_1960_Joseph_IlNuovoCimento.16.6.997}) for general nuclear captures
with RPC in mind should similarly apply for RMC, as shown by Plestid and Hill\,\cite{RMC_2020_Plestid_Hill_arXiv_2010.09509}.
This approximation makes simplifying assumptions about the $e^+e^-$ matrix element by using the energy spectrum of
the on-shell photon and assumes that the virtuality of the $e^+e^-$ pair is small. Following this, the
internal conversion spectrum can be estimated directly from the on-shell photon spectrum, requiring only an on-shell photon
spectrum to predict the total on- and off-shell photon induced background. Plestid and Hill show that this approximation is
most reliable in the high energy region of the positron spectrum, with the next order uncertainty decreasing
as the positron energy approaches the endpoint\,\cite{RMC_2020_Plestid_Hill_arXiv_2010.09509}.

The on-shell RMC photon spectrum was measured by the TRIUMF RMC Spectrometer group on several nuclear targets, including
aluminum\,\cite{PhysRevC.46.1094, Bergbusch_1995, RMC_1998_PhysRevC.58.1767, RMC_1999_PhysRevC.59.2853}.
The experiment was interested in studying the pseudoscalar coupling constant in weak interactions, $g_p$, through
the ratio $g_p/g_a$, where $g_a$ is the axial coupling constant. As such, the focus of the measurements was on the
total rate of RMC, not on a precise model of the high energy tail.

The closure approximation is typically used to describe the
RMC photon energy spectrum, where one assumes the sum of the nuclear final states can be approximated with a single nuclear
transition using the mean excitation energy. This nuclear excitation energy manifests as the closure approximation endpoint, and
is typically considered a free parameter that is fit to data. The closure approximation photon energy spectrum is shown in
Equation\,\ref{eq:closure}, where $x = E_\gamma/k_{\rm max}$ and $k_{\rm max}$ is the spectrum
endpoint\,\cite{CHRISTILLIN1980331}:

\begin{equation}
  \frac{dn}{dx} = \frac{e^2}{\pi}\frac{k_{\rm max}^2}{m_\mu^2}(1 - \frac{N-Z}{A})(1-2x+2x^2)x(1-x)^2 ~.
  \label{eq:closure}
\end{equation}

The TRIUMF RMC Spectrometer group measured a closure approximation endpoint of 90.1 $\pm$ 1.8 MeV and a branching
fraction of $(1.40 \pm 0.11)\times 10^{-5}$ above 57 MeV with respect to ordinary muon capture (OMC) using aluminum
as their nuclear target\,\cite{RMC_1999_PhysRevC.59.2853}.
This endpoint is significantly lower than the kinematic endpoint on aluminum, $\sim$101.9 MeV, as shown in Appendix \ref{rmc_calculation}.
This was the case for all of the nuclear targets -- the fit closure approximation endpoint was $\sim$10 MeV below the
target's kinematic endpoint.
As nothing forbids photon energies up to the kinematic endpoint, there is no reason to expect the spectrum to be 0 between the
measured endpoint and the kinematic endpoint, though it may be suppressed. Predictions using a Fermi gas model by Fearing et al.
\cite{PhysRevC.39.2349, PhysRevC.46.2077} show the photon energy spectrum falling up to the kinematic endpoint, with no tuned
endpoint as found in the closure approximation. A few example spectra from these calculations are shown in Figure\,\ref{fig:Fearing_Spectra}.

\begin{figure}[H]
  \centering
  \includegraphics[width=0.41\linewidth, trim= 0 100 330 0,clip]{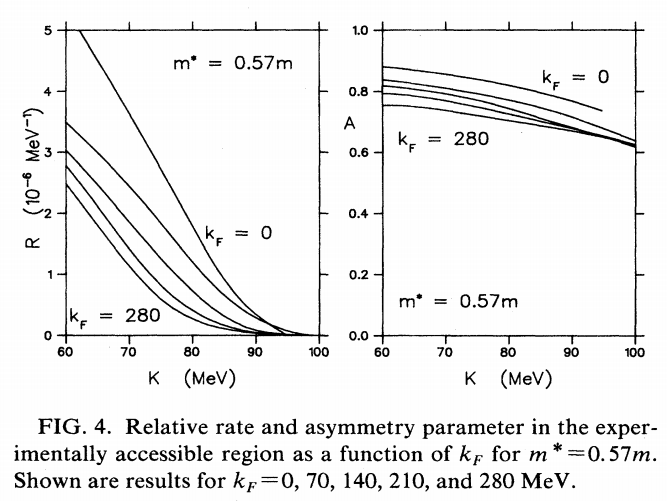}
  \includegraphics[width=0.58\linewidth, trim= 0 250 10 0,clip]{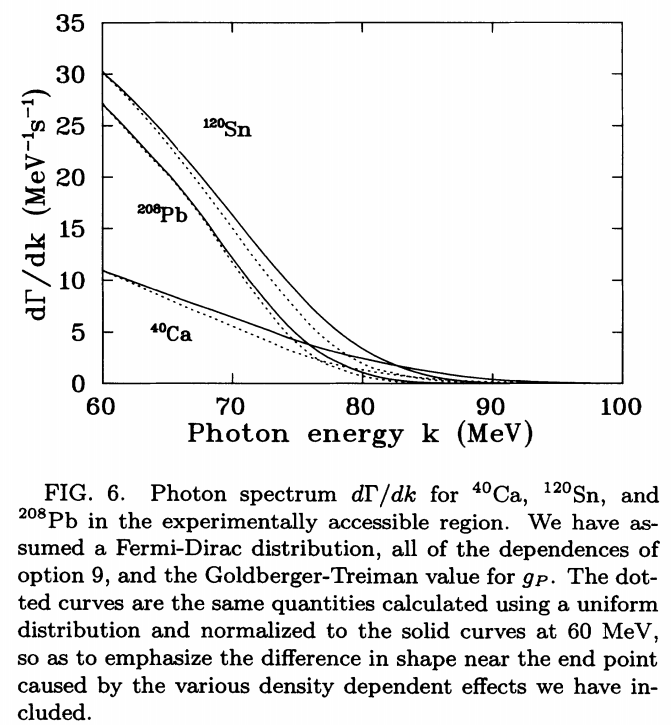}
  \caption{RMC spectra vs $k_{\rm Fermi}$\,\cite{PhysRevC.39.2349} (left) and example RMC spectra\,\cite{PhysRevC.46.2077} (right) as calculated
  using a Fermi gas model.}
  \label{fig:Fearing_Spectra}
\end{figure}

On aluminum, the nuclear target for the currently planned COMET Phase-I and Mu2e searches, the ground state transition energy for the
positron signal is 92.3 MeV, far below the RMC kinematic endpoint of $\sim$101.9 MeV. Unlike the $\MuToEm$ searches,
DIO backgrounds are not a significant background in the positron channel at the sensitivity level
of the current and future experiments,
so the dominant backgrounds are RMC,
RPC, antiprotons,
and cosmic-ray events. The non-RMC backgrounds are expected to contribute similarly in the
positron and electron channel, where the expectation is less than 1 event per experiment for both
searches\,\cite{10.1093/ptep/ptz125, bartoszek2015mu2e}.
Assuming the closure approximation with the measured endpoint, the true positron spectrum would end at $89.6 \pm 1.8$ \mevc\ with a rapidly
falling spectrum, so only resolution and energy loss effects would lead to overlaps with the signals. As the resolution is
approximately 200 \kevc\ for both experiments\,\cite{10.1093/ptep/ptz125, bartoszek2015mu2e} and the two processes are separated by
nearly 3 \mevc, both experiments would
be able to maintain a background expectation of below 1 event per experiment for the measured endpoint.

The dataset on aluminum for the 1999 RMC measurement from TRIUMF only had 3,051 photons above 57 MeV,
so this measurement was not sensitive to photon rates above 90 MeV that are about 3,000
times smaller than the total rate above 57 MeV ($R(E_\gamma >57~{\rm MeV}) = \frac{\Gamma_{RMC}(E_\gamma > 57~{\rm MeV})}{\Gamma_{OMC}}$).
COMET Phase-I and Mu2e will see about $10^{16}$ and $10^{18}$ muon captures respectively\,\cite{10.1093/ptep/ptz125, bartoszek2015mu2e},
and so will have $\sim 10^{11}$ and $10^{13}$ RMC photons above 57 MeV.
To test their sensitivity to RMC beyond the fit endpoint on aluminum, we assume a flat photon energy spectrum tail above 90 MeV up to
the kinematic endpoint with a rate around the sensitivity limit of the 1999 measurement, shown in Equation \ref{eq:rmc_above_90}:

\begin{equation}
  R(E_\gamma > 90~{\rm MeV}) = \frac{1}{3,000} \times R(E_\gamma>57~{\rm MeV}) = 4.7 \times 10^{-9}~~.
  \label{eq:rmc_above_90}
\end{equation}

\noindent
We also make the following simplifying assumptions:
all $e^\pm$ energies from a photon conversion are equally likely, such that the energy sharing distribution
between the $e^+e^-$ pair is flat,
there is a 0.1\% chance of
a photon conversion in the nuclear target, and the tracking efficiency for $\sim$90 \mevc\ positrons is 10\%.
This leads to a positron background rate of $\mathcal{O}(500)$ and $\mathcal{O}(50,000)$ events per \mevc\ at 90 \mevc\ for COMET Phase-I
and Mu2e respectively, before considering internal photon conversions, as shown in Figure \ref{fig:rmc_toymc}.
In this simple model, which is consistent with the existing data on aluminum,
the RMC background estimate would change from below 1 background event to tens of thousands of background events near the signal for
$10^{18}$ muon captures, significantly lessening the discovery potential of $\MuToEp$ signal searches.

\begin{figure}[H]
  \centering
  \includegraphics[width=0.95\linewidth, trim= 0 0 0 0,clip]{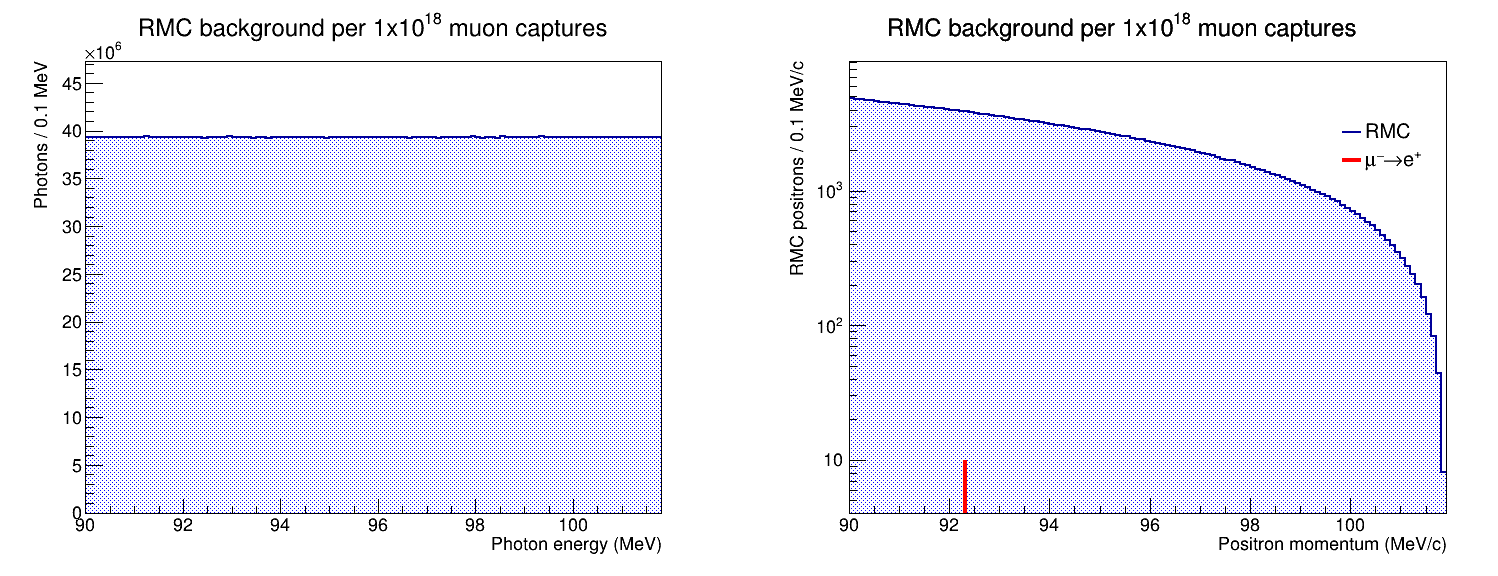}
  \caption{Assumed on-shell RMC photon spectrum high energy tail on aluminum for $10^{18}$ muon captures (left)
    and the corresponding positron spectrum using the simplifying assumptions about the conversion spectrum and
    track reconstruction (right).
    The signal process is also shown using the same tracking efficiency for a sample $\mathcal{R}^{\mu^- e^+}$ value of $10^{-16}$,
    which would likely be discoverable in the absence of an RMC background.
  }
  \label{fig:rmc_toymc}
\end{figure}

\subsection{RMC considerations at future \mupconversion experiments}

The discovery potential of future experiments searching for $\MuToEp$ is entwined with their understanding of the RMC
background, where any claim of new physics must also be able to accurately subtract the SM background.
As the RMC spectrum is not well constrained by existing measurements, muon conversion experiments
should plan to perform RMC measurements on the relevant nuclear targets.
COMET and Mu2e should investigate how to measure the RMC spectrum using the electron and positron spectra,
paying special attention to how to avoid unblinding the $\MuToEp$ signal region of the positron spectrum.
If both the electron and positron track can be reconstructed for an RMC photon conversion, this will allow
one to measure the photon energy while also likely avoiding the $\MuToEp$ signal unblinding. A calorimeter measurement,
using reconstructed photon clusters, would allow a direct measurement of the photon energy spectrum, where the
positron spectrum due to off-shell conversions can be described using the results from Plestid and Hill
given a measurement of the on-shell photon spectrum\,\cite{RMC_2020_Plestid_Hill_arXiv_2010.09509}.
A dedicated experiment to measure the RMC spectrum on the intended nuclear target should also be investigated.

Future muon conversion experiments can also consider choosing a nuclear target whose
$\MuToEp$ ground state transition energy is above the kinematic endpoint of the RMC spectrum, kinematically suppressing this background.
This translates into a requirement that the nuclear mass of N(A,Z-2) is smaller than the nuclear mass of N(A,Z-1).
Several potential target choices were identified by Yeo et al.\,\cite{PhysRevD.96.075027}, where all of the nuclear targets considered had
$\MuToEp$ ground state transition energies greater than the RMC positron kinematic endpoints.
Example targets well suited for $\MuToEp$ measurements are titanium (a common candidate for $\MuToEm$ experiments), sulfur,
and calcium,
which satisfy this mass difference requirement while they also have long enough muon lifetimes to be useful at experiments with delayed data acquisition
to suppress pion backgrounds.
For each nuclear target considered, the authors assumed
a closure approximation with the endpoint set to be the kinematic endpoint and the branching fraction above 57 MeV to be the measured value
by the TRIUMF RMC Spectrometer group.
For an example $10^{18}$ muon stops, they found COMET Phase-II would be able to achieve limits at
90\% CL of $\mathcal{O}(10^{-15})$
for the ground state transition,
a three order of magnitude improvement
upon the current limit from Ref.\,\cite{1998334}.
This is compared to the potential one order of magnitude gain
by using aluminum as the nuclear target\,\cite{PhysRevD.96.075027}, showing how critical
the nuclear target choice can be for the $\MuToEp$ search.

These potential upper limits do not include an assessment of the impact on the limit due to the systematic uncertainty
on the RMC background modeling. The closure approximation does not take into account exclusive transitions to low lying states,
which could lead to a more complicated background spectrum with kinks or knees in the photon energy spectrum. These are likely
to be smoothed by the (internal) pair conversion spectrum, but without understanding the RMC spectrum
it will be difficult to have confidence that an observation of 5-10 events is definitive evidence of $\MuToEp$ conversion, and not instead due
to an RMC transition. An observation of this size is sufficient to claim a discovery at the upcoming $\MuToEm$ searches.
As the goal of these searches is to discover new physics, the experiments must be prepared for the potential discovery of the
LFV and LNV process, \mupconversion, which requires the confident rejection of RMC as an alternate hypothesis.

\section{Concluding remarks: Towards future \mupconversion measurements}
CLFV is a long sought-after signal of physics beyond the SM.
Many experiments have searched for \mueconversion and \mupconversion processes, where the
results are complementary to the best-measured CLFV channel, $\mu \to e \gamma$.
The \mupconversion process is also an LNV channel, which may give insight into the Majorana property of the neutrino.
Although it is not as strong a channel as \znbb is in general, its discovery would provide insight on potential flavor effects
in the neutrino mass generation.
Current and future experiments will have unprecedented sensitivity to both \mueconversion and \mupconversion processes. However, the
search for \mupconversion requires more careful considerations in terms of the nuclear target selection in order
to reach the full discovery potential of these experiments.

RMC is one of the least understood backgrounds at \mueconversion experiments, with the potential to be
the largest background in the \mupconversion search. RMC is theoretically understudied, and previous RMC measurements do not have
the necessary sensitivity in the high energy photon region to sufficiently constrain this background at the currently planned
COMET Phase-I and Mu2e experiments. These experiments will need to measure the RMC photon energy spectrum
in order to confidently reject the SM background in the case of a \mupconversion signal.

An important consideration for the currently planned COMET and Mu2e experiments using aluminum as their nuclear target is
what the next steps will be if they discover the process of \mupconversion. A common idea is to attempt to determine the nature
of the new physics by testing the nuclear target dependence of the conversion
rate\,\cite{Mu2e_II_abusalma2018expression}.
For \mueconversion, Ref.\,\cite{MuToEm_Rates_PhysRevD.66.096002} and \cite{MuToEm_Rates_PhysRevD.80.013002} show for example models
this rate may only vary by $\sim$10-30\% between the models for nuclear targets feasible for COMET Phase-II and Mu2e-II.
The total number of muon captures will likely only be known to $\sim$10\%\,\cite{bartoszek2015mu2e}, uncorrelated between the targets,
making a comparison with less than 15\% uncertainty in the difference between the rates difficult.
It would also require a high statistics discovery to have a precise measurement of the branching fraction, far beyond
the threshold of $\sim$5-10 events needed to claim a discovery for the \mueconversion (and potentially \mupconversion) searches.

In the \mupconversion case, there is more than the ground state transition to consider. Experiments can search for the GDR transition in
addition to the ground-state transition, and measure the relative rate between the ground state and the GDR transition.
As this is a ratio, independent of the number of muon captures,
it is possible this can be better used to test the rate dependence on the nuclear target, helping to understand the
underlying mechanism
for the LNV process. The nuclear dependence of the \mupconversion process needs to be further studied to determine the potential next
steps experiments should take in the case of a \mupconversion discovery.

\appendixstart
\appendix
\section{RMC endpoint calculation on aluminum}
\label{rmc_calculation}

The RMC kinematic endpoint on aluminum can be calculated by considering the ``decay''
of the $\mu^- ~N(A,Z)$ system to $\gamma ~N(A,Z-1) \nu_\mu$ with $p(\nu_\mu) = 0$, which
is then a two body decay with a photon energy given by Equation \ref{eq:rmc_endpoint}.
Since the muon is $\sim$200 times heavier than the electron, and therefore has a much closer orbit in the atom,
the electrons do not participate in the process. The relevant masses are then the muon mass and the incoming and outgoing
nuclei masses, not the atomic masses, where the nuclear mass is given by
$M(A,Z) = M_A - Z\cdot m_e$, where $M_A=A_r\cdot u$, $A_r$ is the relative mass,
u is the atomic mass unit, and $m_e$ is the electron mass.
The final state particles then satisfy $p_{\gamma} = -p_{^{27}Al}$ and the energy of the recoiling magnesium nucleus
is given by:

\begin{equation}
  E_{^{27}Mg} = \frac{M^2 + M_N(^{27}Mg)^2}{2M} = E_{\rm recoil} + M_N(^{27}Mg)~,
  \label{eq:recoil_energy}
\end{equation}

\noindent
where $M$ is the mass of the muonic aluminum system, $M=M(^{27}Al) + m_\mu - B_\mu$.
Table \ref{tab:endpoint_calc_params} shows the input parameters for Equations \ref{eq:recoil_energy} and
\ref{eq:rmc_endpoint}. The resulting kinematic RMC endpoint on aluminum is 101.867 MeV. 
The corresponding positron energy endpoint that is relevant for \mupconversion experiments is one electron mass below this,
which is listed in Table\,\ref{table_nuclei_list} for some example nuclei.

\begin{specialtable}[H]
  \centering
  \caption{Parameters used in the RMC endpoint energy calculation on aluminum.}
  \label{tab:endpoint_calc_params}
    \begin{tabular}{cr@{.}llc}
      \toprule
      \textbf{Parameter} & \multicolumn{3}{c}{\textbf{Value}} & \textbf{Reference}\\
      \midrule
      $m_\mu$         & 105&6583745&\mevc$^2$&\cite{Nuclear_Data_PhysConstants}\\
      $m_e$          & 0&5109989461&\mevc$^2$&\cite{Nuclear_Data_PhysConstants}\\
      $1u$           & 931&49410242&\mevc$^2$&\cite{Nuclear_Data_PhysConstants}\\
      $B_\mu$         & 0&464&MeV&\cite{PhysRevD.84.013006}\\
      $A_r(^{27}Al)$  &  26&98153841&u&\cite{Wang_2017}\\
      $M_N(^{27}Al)$  &  25126&501&\mevc$^2$&  \\
      $A_r(^{27}Mg)$  &  26&98434063&u&\cite{Wang_2017}\\
      $M_N(^{27}Mg)$  &  25129&622&\mevc$^2$&  \\
      $E_{\rm recoil}$  &  0&206&\mevc$^2$ &  \\
      \bottomrule
    \end{tabular}
\end{specialtable}

\begin{specialtable}[H]
  \centering
\caption{The energy of the signal positron ($E_{\mu^-e^+}$) obtained from Equation\,\ref{eq:mup_energy},
the RMC positron endpoint energy ($E^{\rm{end}}_{\rm{RMC}}$) of the ground state transition obtained from Equation\,\ref{eq:rmc_endpoint},
and the lifetime of the muonic atom ($\tau_{\mu^-}$)
of some nuclei considered for \mupconversion experiments. The energy of the signal electron ($E_{\mu^-e^-}$) for
\mueconversion are also shown for comparison.
Nuclear masses required for these calculations are taken from AME2016 data\,\cite{Wang_2017}.
The lifetime data is from Ref.\,\cite{PhysRevC.35.2212}.
\label{table_nuclei_list}}
\begin{tabular}{ccccc}
\toprule
\textbf{Nuclide} &
\boldmath $E_{\mu^-e^+}$ \unboldmath \textbf{[MeV]} &
\boldmath $E^{\rm{end}}_{\rm{RMC}}$ \unboldmath \textbf{[MeV]}  &
\boldmath $\tau_{\mu^-}$ \unboldmath \textbf{[ns]} &
\boldmath $E_{\mu^-e^-}$ \unboldmath \textbf{[MeV]} \\
\midrule
${}^{27}{\rm Al}$ & 92.30   & 101.36 & 864 & 104.97 \\
${}^{32}{\rm S}$  & 101.80  & 102.03 & 555 & 104.76 \\
${}^{40}{\rm Ca}$ & 103.55  & 102.06 & 333 & 104.39 \\
${}^{48}{\rm Ti}$ & 98.89   & 99.17  & 329 & 104.18 \\
\bottomrule
\end{tabular}
\end{specialtable}

\vspace{6pt}

\authorcontributions{Writing---original draft preparation, M. L. and M. M.; writing---review and editing, M.L. and M. M. All authors have read and agreed to the published version of the manuscript.}

\funding{This research of M.L. was supported by Institute for Basic Science (IBS-R017-D1-2021-a00) of Korea.
  This research of M.M. was in part funded by the ``Research in the Energy, Cosmic and Intensity Frontiers at Northwestern University''
  award, DE-SC0015910.}

\institutionalreview{Not applicable.}

\informedconsent{Not applicable.}

\dataavailability{The data used in this study can be provided by the authors upon request.}

\acknowledgments{The authors M.L. and M.M. are collaborators of the COMET and Mu2e collaborations, respectively. 
The author M.L. thanks KEK and J-PARC, Japan for their support of infrastructure and the operation of COMET. 
This work of M.L. is supported in part by:
Japan Society for the Promotion of Science (JSPS) KAKENHI Grant Nos. 25000004 and 18H05231; JSPS KAKENHI Grant No.JP17H06135;
Belarusian Republican Foundation for Fundamental Research Grant F18R-006;
National Natural Science Foundation of China (NSFC) under Contracts No. 11335009 and 11475208;
Research program of the Institute of High Energy Physics (IHEP) under Contract No. Y3545111U2;
the State Key Laboratory of Particle Detection and Electronics of IHEP, China, under Contract No.H929420BTD;
Supercomputer funding in Sun Yat-Sen University, China;
National Institute of Nuclear Physics and Particle Physics (IN2P3), France;
Shota Rustaveli National Science Foundation of Georgia (SRNSFG), grant No. DI-18-293;
Deutsche Forschungsgemeinschaft grant STO 876/7-1 of Germany;
Joint Institute for Nuclear Research (JINR), project COMET \#1134;
Institute for Basic Science (IBS) of Republic of Korea under Project No. IBS-R017-D1-2021-a00;
Ministry of Education and Science of the Russian Federation and by the Russian Fund for Basic Research grants: 17-02-01073, 18-52-00004;
Science and Technology Facilities Council, United Kingdom;
JSPS London Short Term Predoctoral Fellowship program, Daiwa Anglo-Japanese Foundation Small Grant; and Royal Society International Joint Projects Grant.
The author M.M. is grateful for the vital contributions of the
Fermilab staff and the technical staff of the participating institutions to Mu2e. This work was supported by the US Department of Energy;
the Istituto Nazionale di Fisica Nucleare, Italy; the Science and Technology Facilities Council, UK;
the Ministry of Education and Science, Russian Federation; the National Science Foundation, USA; the Thousand Talents Plan, China;
the Helmholtz Association, Germany; and the EU Horizon 2020 Research and Innovation Program under the Marie Sklodowska-Curie Grant Agreement
No. 690835, 734303, 822185, 858199.
This document was prepared in part by members of the Mu2e Collaboration using the resources of the Fermi National Accelerator Laboratory (Fermilab),
a U.S. Department of Energy, Office of Science, HEP User Facility. Fermilab is managed by Fermi Research Alliance, LLC (FRA), acting under Contract
No. DE-AC02-07CH11359.}

\conflictsofinterest{The authors declare no conflict of interest.}

\end{paracol}
\reftitle{References}

\end{document}